\documentclass[12pt]{article}

\usepackage{graphicx}
\usepackage{longtable}

\begin{document}

\title{Theoretical Study on Rotational Bands and Shape
Coexistence of $^{183,185,187}${Tl} in the Particle Triaxial-Rotor
Model}

\author{{Guo-Jie Chen$^{1,}$\footnote{Email: chengj126@126.com },
Yu-xin Liu$^{2,3,4,5,}$\footnote{Corresponding author, Email
address: liuyx@phy.pku.edu.cn} , Hui-chao Song$^{6}$, and Hui Cao$^{1}$ }\\[3mm]
\normalsize{$^1$ School of Science, Foshan University, Foshan
528000, China}\\[-1mm]
\normalsize{$^2$ Department of Physics, Peking University, Beijing
100871, China}\\[-1mm]
\normalsize{$^3$ The Key Laboratory of Heavy Ion Physics,
Ministry of Education, Beijing 100871, China}\\[-1mm]
\normalsize{$^4$ Institute of Theoretical Physics, Academia
Sinica, Beijing 100080, China}\\[-1mm]
\normalsize{$^5$ Center of Theoretical Nuclear Physics, National
Laboratory of Heavy Ion Accelerator, } \\[-1mm]
\normalsize{Lanzhou 730000, China} \\[-1mm]
\normalsize{$^6$ Department of Physics, Ohio State University,
Columbus, OH43210, USA} }

\date{\today}

\maketitle

%
%
%
%
%

\begin{abstract}
By taking the particle triaxial-rotor model with variable moment of
inertia, we investigate the energy spectra, the deformations and the
single particle configurations of the nuclei $^{183,185,187}$Tl
systemically. The calculated energy spectra agree with experimental
data quite well. The obtained results indicate that the aligned
bands observed in $^{183,185,187}$Tl originate from the
$[530]{\frac{1}{2}}^{-}$, $[532]{\frac{3}{2}}^{-}$,
$[660]{\frac{1}{2}}^{+}$ proton configuration coupled to a prolate
deformed core, respectively. Whereas, the negative parity bands
built upon the ${\frac{9}{2}}^{-}$ isomeric states in
$^{183,185,187}$Tl are formed by a proton with the
$[505]{\frac{9}{2}}^{-}$ configuration coupled to a core with
triaxial oblate deformation, and the positive parity band on the
${\frac{13}{2}}^{+}$ isomeric state in $^{187}$Tl is generated by a
proton with configuration $[606]{\frac{13}{2}}^{+}$ coupled to a
triaxial oblate core.
\end{abstract}




\newpage

\section{Introduction}

It has been known that nuclei in the $Z=82$ region are rich in shape
coexistence. In particular, the important deformation driving
orbitals has been assigned as the $h_{9/2}$ and $ i_{13/2}$ proton
shells \cite{Heyde83,Wood92}. In odd-mass Tl isotopes (with $Z=81$),
one-particle--two-hole (1p-2h) intruder states and shape coexistence
have been discovered through the observation of low-lying $9/2^{-}$
isomeric states \cite{DS63}. The structure of these isomeric states
was confirmed to be decided by the odd proton occupying the
$h_{9/2}$ intruder orbital \cite{Newton70,BBJ85}. Later, the
rotational bands associated with both oblate ($\pi $h$_{9/2}$, $\pi
$i$_{13/2})$ and prolate ($\pi $h$_{9/2}$ , $\pi $i$_{13/2}$, $\pi
$f$_{7/2})$ structures have been observed in lighter isotopes
$^{185,187}$Tl \cite{Lane95}. The band-head of the 1p-2h oblate $\pi
$h$_{9/2}$ intruder band has been observed to lie lowest in energy
near $N$=108. In contrast, the band-head of the prolate intruder
band based on the $i_{13/2}$ structure has been predicted to
decrease continuously in excitation energy as the neutron number
decreases beyond the neutron mid-shell. This prolate structure is
presumably formed by coupling the odd $i_{13/2}$ proton to the
prolate Hg core with 4p-6h structure \cite{Lane95}. Recently, a
rotational-like yrast cascade was established in $^{183}$Tl and
assigned to associate with the prolate $i_{13/2}$
structure \cite{Reviol00}. Furthermore the band-head energy of its
yrast band was later determined \cite{Muiku01}.

Besides the coexistence of prolate and oblate shapes mentioned
above, the signature splitting observed in the $[505]9/2^{-}$ band
in $^{187}$Tl which is significantly larger than that observed in
its heavier odd-mass isotopes with $A \ge 191$ suggests that there
may exist triaxial deformation \cite{Heyde83,Lane95} and the
discrepancy between the calculated equilibrium energy and the
experimental data of the band-head energy of the
$[606]{\frac{13}{2}}^{+}$ band in $^{187}$Tl hints that there may
also involve triaxial deformation \cite{Lane95}. However, no concrete
investigations on the triaxiality in $^{185,187}$Tl have been
reported up to now. Furthermore, there does not exist, at present,
a systematic theoretical investigation on the structure of $^{183}$Tl.
In addition, whether the $[532]{\frac{3}{2}}^{-}$ ($h_{9/2}$) state
in $^{185,187}$Tl can be distinguished from the
$[530]{\frac{1}{2}}^{-}$ ($f_{7/2}$) state (the band originated from
such a configuration has not yet been observed in $^{185}$Tl) has
not yet been determined definitely \cite{Lane95}.

On the theoretical side, it has been well established that the total
energy surface calculation is quite successful in studying the
equilibrium shape of a nucleus and shape coexistence (see for
example Refs. \cite{Bengtsson,Xu,Frauend}). In addition, the projected
shell model \cite{HaraSun} and particle triaxial-rotor
model \cite{Meyer745,Larsson78,Ring80} are also suitable to study
triaxial deformation and configuration mixing \cite{Sheikh,Hamamoto}.
However, the triaxial deformation in the light odd-{\it A} Tl-isotopes has
not yet been studied. Because of its simplicity, we take the
particle triaxial-rotor model with variable moment of inertia of
the core to analyze the structure and deformation of the energy bands in
$^{183,185,187}$Tl systemically and to identify their microscopic
configuration.

The paper is organized as follows. After this introduction, we
describe briefly the formalism of the particle triaxial-rotor model
in Section II. In Section III, we describe our calculation and obtained
results. In Section IV, we give a summary and brief remark.

\section{Particle Triaxial-Rotor Model }

In the particle rotor model, the Hamiltonian of an odd-{\it A} nucleus
is usually written as \cite{Meyer745,Larsson78,Ring80}
\begin{equation}
\label{eq1} \hat {H} = \hat {H}_{core} + \hat {H}_{s.p.} + \hat
{H}{ }_{pair} \quad .
\end{equation}

In the case of triaxial deformation, the Hamiltonian of the
even-even core is given as
\begin{equation}
\label{eq2}
\hat {H}_{core} = \sum\limits_{i = 1}^3 {\frac{\hbar ^2R_i ^2}{2\Im _i }} =
\sum\limits_{i = 1}^3 {\frac{\hbar ^2(I_i - j_i )^2}{2\Im _i }} ,
\end{equation}

\noindent
where $R$, $I$ and $j$ are the angular momentum of the core, the nucleus and the
single particle, respectively. The three rotational moments of inertia are
assumed to be connected by a relation of hydrodynamical type
\begin{equation}
\label{eq3} \Im _\kappa = \frac{4}{3}\Im _0 (I) \sin ^2(\gamma +
\frac{2\pi }{3}\kappa ) \quad ,
\end{equation}

\noindent
with
\begin{equation}
\label{eq4}
\Im _0 (I) = \Im _0 \sqrt {1 + bI(I + 1)}
\end{equation}

\noindent being the variable moment of inertia \cite{MSB69} of the
core to replace the original constant $\Im _0$ to improve the
calculation. In present calculation, we take $b = 0.013$ as the
same as that in Refs. \cite{ZS02,SLZ04,CSL05}.

$\hat {H}_{s.p.} $ describes the Hamiltonian of the unpaired
single particle. In the triaxial deformed field of the even-even
core , $\hat {H}_{s.p.} $ is given by
\begin{eqnarray} \label{eq5}
\hat {H}_{s.p.}& = &- \frac{\hbar ^2}{2m}\nabla ^2 +
\frac{1}{2}m\omega _0^2 \{1 - 2\beta [Y_{20} \cos \gamma +
\frac{1}{\sqrt 2 }(Y_{22} + Y_{2 - 2} )\sin \gamma ]\} \nonumber \\
& & - \kappa \hbar \omega _0 \{2 l \cdot s + \mu (l^2 - < l_N >
^2)\} \, ,
\end{eqnarray}

\noindent where $\kappa $ and $\mu $ are Nilsson parameters,
$Y_{2q} $ is the rank-2 spherical harmonic function.

$\hat {H}_{pair}$ is the Hamiltonian to represent the pairing
correlation which can be treated in the Bardeen-Cooper-Schrieffer
(BCS) formalism.

The single-particle wavefunction can be expressed as
\begin{equation}
\label{eq6}
\left| \nu \right\rangle = \sum\limits_{Nlj\Omega } {C_{Nlj\Omega }^{(\nu )}
\left| {Nlj\Omega } \right\rangle } \quad ,
\end{equation}

\noindent where $\nu $ is the sequence number of the single-particle
orbitals, $\vert N l j \Omega \rangle $ represents the corresponding
Nilsson state, $C_{Nlj\Omega }^{(\nu )} $ is the coefficient to
identify the configuration mixing. Diagonalizing the single-particle
Hamiltonian in the basis $\left| {Nlj\Omega } \right\rangle $, we
can obtain the $C^{(\nu )}_{Nlj\Omega} $ and the single-particle
eigenvalue $\varepsilon _\nu $. The corresponding quasi-particle
energy can then be determined by $E_{\nu} = \sqrt {(\varepsilon _\nu
- \lambda )^2 + \Delta ^2} $, with $\lambda $ and $\Delta $ being
the Fermi energy and the energy gap, respectively.

The total Hamiltonian in Eq.(\ref{eq1}) can be diagonalized in the
symmetrically strong coupling basis
\begin{equation}
\label{eq7} \left| {IKM\nu } \right\rangle = \sqrt {\frac{2I +
1}{16\pi ^2}} \left[ D_{MK}^{I} \alpha _{\nu} ^{\dag} \left|
\tilde {0} \right\rangle + (-1)^{I - K}D_{M - K}^{I} \alpha
_{\tilde {\nu }}^{\dag} \left| \tilde {0} \right\rangle  \right]
\, ,
\end{equation}

\noindent where $\alpha _\nu ^ + $ is creation operator of the
single nucleon (in present case, proton) in the orbital $\vert \nu
\rangle $, $D_{MK}^I $ is the rotational matrix.

\section{Calculation and Results}

In the present calculation to investigate the property of
$^{183,185,187}$Tl, we take the $\kappa $ and $\mu $ in standard
values \cite{BR85}, i.e., 0.054, 0.690, respectively, and the pairing
gap parameter as $\Delta = 12 / \sqrt A $. To improve the agreement
between calculated results and experimental data, we introduce a
Coriolis attenuation factor {$\xi $} and take value as that giving
the best agreement between the calculated and experimental energy
spectra. We found that, when {$\xi =0.95$}, the calculated results
agree best with the experimental data of $^{183,185,187}$Tl. In
general principle, in order to describe the nuclear property more
accurately and to make better agreement between calculated and
experimental data, it is necessary to involve sufficient
single-particle orbitals near the Fermi surface in the calculation.
Then we take 13 orbitals near the Fermi surface to couple with the
core for $^{183}$Tl, $^{185}$Tl, $^{187}$Tl, respectively. Practical
calculation shows that the Fermi levels of the bands 6 (we denote
the band labels here as the same as those for nucleus $^{187}$Tl in
Ref.~\cite{Lane95}, so that the similar bands can be compared) of
the nuclei lie between the 20th and the 21st single particle
orbitals, and the others lie between the 19th and the 20th. For the
deformation parameters {$\beta $} and $\gamma$ of $^{185,187}$Tl, we
take those given in Ref.\cite{Lane95} as the trial initial values to
fit. For the deformation parameters of $^{183}$Tl, since there does
not exist any report to discuss them, we take the values of its
neighbor nucleus $^{185}$Tl \cite{Lane95} as the trial initial ones.
Then we accomplished a series diagonalization of the total
Hamiltonian with various values of $\beta $ and $\gamma$ to make the
calculation error $\chi ^2 = \frac{1}{N} \sum _{j} (E^{cal}_{j} -
E^{exp}_{j} )^2 $ of the spectrum of a band (where $N$ is the number
of levels in the band) smaller (in such a process, the band-head
energy is fixed artificially with the definite angular momentum
assigned in experiment. The best fit is, in fact, focused on the
energy separations). Meanwhile, it should be noted that the
parameter sector we used in the present work is the same as that
taken in the book by Nilsson and Ragnarsson \cite{NR95}, where the
value of $\beta$ can be positive or negative, the value of $\gamma$
varies from 0 to 30 degrees. The best fitted values of the $\beta$
and $\gamma$ are listed in Table 1, 2, 3 for nucleus $^{183}$Tl,
$^{185}$Tl, $^{187}$Tl, respectively. At the same time, we obtain the
total wavefunctions in terms of the single-particle orbitals which,
as mentioned above, is fixed by diagonalizing the single particle
Hamiltonian at each set of deformation parameters ($\beta$,
$\gamma$). The calculated main components of the single-particle
orbitals in terms of the Nilsson levels (in the case of best fitted
deformation parameters) of nucleus $^{183}$Tl, $^{185}$Tl,
$^{187}$Tl are also listed in Table 1, 2, 3, respectively. The
resulting energy spectra for the nuclei $^{183,185,187}$Tl,
as obtained from the best fit, are illustrated in Fig.1.
From inspecting the results shown in Fig.~1, we observe a good agreement
with the experimental data.

\begin{table}[htbp]
\begin{footnotesize}
\caption{The deformation parameters and the main components of the
single-particle levels $\vert \nu \rangle $ near the Fermi surface
in terms of the Nilsson levels of the bands in $^{183}$Tl (the
initial values of the deformation parameters are taken as those of
$^{185}$Tl in Ref.~\cite{Lane95}.) } \vspace*{-2mm}
\begin{center}
\begin{tabular}{|c|c|c|c|c|l|}
\hline {} & \multicolumn{2}{|c|}{$\beta$} &
\multicolumn{2}{|c|}{$\gamma$} & {} \\ \cline{2-5} band & initial
& fitted & initial & fitted & $\nu \rangle \;\; $  wave function
in terms of $\vert N l j \Omega \rangle $ \\[-2mm]
{} & value & value & value & value & {} \\ \hline
 {} & {} & {} & {} & {} & $\vert 19 \rangle \; \; 0.856 \vert 5h_{11/2}
      \frac{1}{2} \rangle + 0.425 \vert 5 f_{5/2} \frac{1}{2} \rangle
     -0.182 \vert 5 f_{7/2} \frac{1}{2} \rangle $ \\
 {} & {} & {} & {} & {} & $\vert 20 \rangle \; \; 0.986 \vert 5h_{9/2}
  \frac{9}{2} \rangle + 0.103 \vert 5 h_{9/2} \frac{5}{2} \rangle $ \\
 band 3 & $-0.162$ & $-0.168$ & 0 & $15^{\circ}$ & $\vert 21 \rangle \; \;
     0.764 \vert 5h_{9/2} \frac{7}{2} \rangle + 0.601 \vert 5 f_{7/2}
     \frac{7}{2} \rangle -0.169 \vert 5 h_{9/2} \frac{3}{2} \rangle $ \\
($[505]\frac{9}{2}^{-}$) & {} & {} & {} & {} & $\vert 22 \rangle
     \;\; 0.819 \vert 5h_{9/2}\frac{5}{2} \rangle + 0.385 \vert 5 h_{9/2}
      \frac{1}{2} \rangle -0.210 \vert 5 f_{5/2} \frac{5}{2} \rangle $ \\
 {} & {} & {} & {} & {} & $\vert 23 \rangle \; \; 0.733 \vert 5h_{9/2}
      \frac{3}{2} \rangle + 0.392 \vert 5 h_{9/2} \frac{1}{2} \rangle
     +0.378 \vert 5 h_{9/2} \frac{5}{2} \rangle $ \\               \hline
 {} & {} & {} & {} & {} & $\vert 19 \rangle \; \; 0.995 \vert 4g_{9/2}
      \frac{7}{2} \rangle $ \\
 {} & {} & {} & {} & {} & $\vert 20 \rangle \; \; 0.967 \vert 4d_{5/2}
  \frac{5}{2} \rangle + 0.217 \vert 4 g_{7/2} \frac{5}{2} \rangle
  + 0.130 \vert 4 g_{9/2}\frac{5}{2} \rangle $ \\
 band 6 & 0.267 & 0.270 & 0 & 0 & $\vert 21 \rangle \; \;
     0.951 \vert 6i_{13/2} \frac{1}{2} \rangle + 0.545 \vert 6 g_{9/2}
     \frac{1}{2} \rangle $ \\
($[660]\frac{1}{2}^{+}$) & {} & {} & {} & {} & $\vert 22 \rangle
     \;\; 0.945 \vert 6i_{13/2}\frac{3}{2} \rangle + 0.511 \vert 6 g_{9/2}
      \frac{3}{2} \rangle $ \\
 {} & {} & {} & {} & {} & $\vert 23 \rangle \; \; 0.902 \vert 4d_{3/2}
      \frac{3}{2} \rangle + 0.251 \vert 4 s_{1/2} \frac{1}{2} \rangle
     +0.216 \vert 5 d_{5/2} \frac{3}{2} \rangle $ \\  \hline
\end{tabular}
\label{tab1}
\end{center}
\end{footnotesize}
\end{table}

\begin{table}[htbp]
\begin{footnotesize}
\caption{The deformation parameters and the main components of the
single-particle levels $\vert \nu \rangle $ near the Fermi surface
in terms of the Nilsson levels of the bands in $^{185}$Tl (the
initial values of the deformation parameters are taken from
Ref.~\cite{Lane95}.) }  \vspace*{-2mm}
\begin{center}
\begin{tabular}{|c|c|c|c|c|l|}
\hline {} & \multicolumn{2}{|c|}{$\beta$} &
\multicolumn{2}{|c|}{$\gamma$} & {} \\ \cline{2-5} band & initial
& fitted & initial & fitted & $\nu \rangle \;\; $  wave function
in terms of $\vert N l j \Omega \rangle $ \\[-2mm]
{} & value & value & value & value & {} \\ \hline
 {} & {} & {} & {} & {} & $\vert 18 \rangle \; \; 0.993 \vert 5h_{11/2}
      \frac{9}{2} \rangle $ \\
 {} & {} & {} & {} & {} & $\vert 19 \rangle \; \; 0.872 \vert 5h_{9/2}
  \frac{1}{2} \rangle - 0.420 \vert 5 f_{5/2} \frac{1}{2} \rangle
  + 0.183 \vert 5 f_{7/2} \frac{1}{2} \rangle $ \\
 band 2 & 0.245 & 0.247 & 0 & $0$ & $\vert 20 \rangle \; \;
     0.910 \vert 5h_{9/2} \frac{3}{2} \rangle + 0.335 \vert 5 f_{5/2}
     \frac{3}{2} \rangle +0.195 \vert 5 f_{7/2} \frac{3}{2} \rangle $ \\
($[532]\frac{3}{2}^{-}$) & {} & {} & {} & {} & $\vert 21 \rangle
     \;\; 0.997 \vert 5h_{11/2}\frac{11}{2} \rangle $ \\
 {} & {} & {} & {} & {} & $\vert 22 \rangle \; \; 0.764 \vert 5f_{7/2}
      \frac{1}{2} \rangle + 0.475 \vert 5 p_{3/2} \frac{1}{2} \rangle
     +0.321 \vert 5 h_{9/2} \frac{1}{2} \rangle $ \\               \hline
 {} & {} & {} & {} & {} & $\vert 19 \rangle \; \; 0.681 \vert 5h_{11/2}
      \frac{1}{2} \rangle + 0.449 \vert 5 f_{5/2} \frac{1}{2} \rangle
     -0.352 \vert 5 f_{7/2} \frac{1}{2} \rangle $ \\
 {} & {} & {} & {} & {} & $\vert 20 \rangle \; \; 0.976 \vert 5h_{9/2}
  \frac{9}{2} \rangle + 0.121 \vert 5 h_{9/2} \frac{5}{2} \rangle $ \\
 band 3 & $-0.162$ & $-0.164$ & 0 & $15^{\circ}$ & $\vert 21 \rangle \; \;
     0.770 \vert 5h_{9/2} \frac{9}{2} \rangle + 0.332 \vert 5 f_{7/2}
     \frac{7}{2} \rangle -0.182 \vert 5 h_{9/2} \frac{3}{2} \rangle $ \\
($[505]\frac{9}{2}^{-}$) & {} & {} & {} & {} & $\vert 22 \rangle
     \;\; 0.815 \vert 5h_{9/2}\frac{5}{2} \rangle + 0.394 \vert 5 h_{9/2}
      \frac{1}{2} \rangle -0.221 \vert 5 f_{5/2} \frac{5}{2} \rangle $ \\
 {} & {} & {} & {} & {} & $\vert 23 \rangle \; \; 0.752 \vert 5h_{9/2}
      \frac{3}{2} \rangle - 0.392 \vert 5 h_{9/2} \frac{1}{2} \rangle
     -0.372 \vert 5 h_{9/2} \frac{5}{2} \rangle $ \\               \hline
 {} & {} & {} & {} & {} & $\vert 19 \rangle \; \; 0.998 \vert 4g_{9/2}
      \frac{7}{2} \rangle $ \\
 {} & {} & {} & {} & {} & $\vert 20 \rangle \; \; 0.935 \vert 4d_{5/2}
  \frac{5}{2} \rangle + 0.235 \vert 4 g_{7/2} \frac{5}{2} \rangle
  + 0.151 \vert 4 g_{9/2}\frac{5}{2} \rangle $ \\
 band 6 & 0.267 & 0.268 & 0 & 0 & $\vert 21 \rangle \; \;
     0.941 \vert 6i_{13/2} \frac{1}{2} \rangle + 0.337 \vert 6 g_{9/2}
     \frac{1}{2} \rangle $ \\
($[660]\frac{1}{2}^{+}$) & {} & {} & {} & {} & $\vert 22 \rangle
     \;\; 0.952 \vert 6i_{13/2}\frac{3}{2} \rangle + 0.298 \vert 6 g_{9/2}
      \frac{3}{2} \rangle $ \\
 {} & {} & {} & {} & {} & $\vert 23 \rangle \; \; 0.921 \vert 4d_{3/2}
      \frac{3}{2} \rangle + 0.238 \vert 4 s_{1/2} \frac{1}{2} \rangle
     +0.208 \vert 5 d_{5/2} \frac{3}{2} \rangle $ \\  \hline
\end{tabular}
\label{tab2}
\end{center}
\end{footnotesize}
\end{table}

\begin{table}[htbp]
\begin{footnotesize}
\caption{The deformation parameters and the main components of the
single-particle levels $\vert \nu \rangle $ near the Fermi surface
in terms of the Nilsson levels of the bands in $^{187}$Tl (the
initial values of the deformation parameters are taken from
Ref.~\cite{Lane95}.) }   \vspace*{-2mm}
\begin{center}
\tabcolsep=4.5pt
\begin{tabular}{|c|c|c|c|c|l|}
\hline {} & \multicolumn{2}{|c|}{$\beta$} &
\multicolumn{2}{|c|}{$\gamma$} & {} \\ \cline{2-5} band & initial
& fitted & initial & fitted & $\nu \rangle \;\; $  wave function
in terms of $\vert N l j \Omega \rangle $ \\[-2mm]
{} & value & value & value & value & {} \\ \hline
 {} & {} & {} & {} & {} & $\vert 18 \rangle \; \; 0.997 \vert 5h_{11/2}
      \frac{9}{2} \rangle $ \\
 {} & {} & {} & {} & {} & $\vert 19 \rangle \; \; 0.849 \vert 5h_{9/2}
  \frac{1}{2} \rangle + 0.429 \vert 5 f_{5/2} \frac{1}{2} \rangle
  - 0.169 \vert 5 f_{7/2} \frac{1}{2} \rangle $ \\
 band 1 & 0.250 & 0.253 & 0 & $0$ & $\vert 20 \rangle \; \;
     0.909 \vert 5h_{9/2} \frac{3}{2} \rangle + 0.325 \vert 5 f_{5/2}
     \frac{1}{2} \rangle +0.191 \vert 5 f_{7/2} \frac{3}{2} \rangle $ \\
($[530]\frac{1}{2}^{-}$) & {} & {} & {} & {} & $\vert 21 \rangle
     \;\; 1.000 \vert 5h_{11/2}\frac{11}{2} \rangle $ \\
 {} & {} & {} & {} & {} & $\vert 22 \rangle \; \; 0.741 \vert 5f_{7/2}
      \frac{1}{2} \rangle + 0.480 \vert 5 p_{3/2} \frac{1}{2} \rangle
     +0.327 \vert 5 h_{9/2} \frac{1}{2} \rangle $ \\               \hline
 {} & {} & {} & {} & {} & $\vert 18 \rangle \; \; 0.997 \vert 5h_{11/2}
      \frac{9}{2} \rangle $ \\
 {} & {} & {} & {} & {} & $\vert 19 \rangle \; \; 0.868 \vert 5h_{9/2}
  \frac{1}{2} \rangle + 0.410 \vert 5 f_{5/2} \frac{1}{2} \rangle
  - 0.162 \vert 5 f_{7/2} \frac{1}{2} \rangle $ \\
 band 2 & 0.234 & 0.237 & 0 & $0$ & $\vert 20 \rangle \; \;
     0.916 \vert 5h_{9/2} \frac{3}{2} \rangle + 0.312 \vert 5 f_{5/2}
     \frac{3}{2} \rangle +0.189 \vert 5 f_{7/2} \frac{3}{2} \rangle $ \\
($[532]\frac{3}{2}^{-}$) & {} & {} & {} & {} & $\vert 21 \rangle
      \;\; 1.000 \vert 5h_{11/2}\frac{11}{2} \rangle $ \\
 {} & {} & {} & {} & {} & $\vert 22 \rangle \; \; 0.766 \vert 5f_{7/2}
      \frac{1}{2} \rangle + 0.469 \vert 5 p_{3/2} \frac{1}{2} \rangle
     +0.300 \vert 5 h_{9/2} \frac{1}{2} \rangle $ \\               \hline
 {} & {} & {} & {} & {} & $\vert 19 \rangle \; \; 0.695 \vert 5h_{11/2}
      \frac{1}{2} \rangle + 0.447 \vert 5 f_{5/2} \frac{1}{2} \rangle
     -0.360 \vert 5 f_{7/2} \frac{1}{2} \rangle $ \\
 {} & {} & {} & {} & {} & $\vert 20 \rangle \; \; 0.986 \vert 5h_{9/2}
  \frac{9}{2} \rangle - 0.101 \vert 5 h_{9/2} \frac{5}{2} \rangle $ \\
 band 3 & $-0.162$ & $-0.162$ & 0 & $15^{\circ}$ & $\vert 21 \rangle \; \;
     0.780 \vert 5h_{9/2} \frac{9}{2} \rangle + 0.348 \vert 5 f_{7/2}
     \frac{7}{2} \rangle -0.169 \vert 5 h_{9/2} \frac{3}{2} \rangle $ \\
($[505]\frac{9}{2}^{-}$) & {} & {} & {} & {} & $\vert 22 \rangle
     \;\; 0.825 \vert 5h_{9/2}\frac{5}{2} \rangle + 0.382 \vert 5 h_{9/2}
      \frac{1}{2} \rangle -0.201 \vert 5 f_{5/2} \frac{5}{2} \rangle $ \\
 {} & {} & {} & {} & {} & $\vert 23 \rangle \; \; 0.741 \vert 5h_{9/2}
      \frac{3}{2} \rangle - 0.386 \vert 5 h_{9/2} \frac{1}{2} \rangle
     -0.384 \vert 5 h_{9/2} \frac{5}{2} \rangle $ \\               \hline
 {} & {} & {} & {} & {} & $\vert 23 \rangle \; \; 0.997 \vert 6i_{13/2}
      \frac{13}{2} \rangle $ \\
 {} & {} & {} & {} & {} & $\vert 24 \rangle \; \; 0.991 \vert 6i_{13/2}
  \frac{11}{2} \rangle $ \\
 band 5 & $-0.189$ & $-0.192$ & 0 & $11.3^{\circ}$ & $\vert 25 \rangle \; \;
     0.974 \vert 6i_{13/2} \frac{9}{2} \rangle  $ \\
($[606]\frac{13}{2}^{+}$) & {} & {} & {} & {} & $\vert 26 \rangle
      \;\; 0.931 \vert 6i_{13/2}\frac{7}{2} \rangle + 0.298 \vert 6 i_{13/2}
      \frac{3}{2} \rangle $ \\
 {} & {} & {} & {} & {} & $\vert 27 \rangle \; \; 0.805 \vert 6i_{13/2}
      \frac{5}{2} \rangle + 0.471 \vert 6 i_{13/2} \frac{1}{2} \rangle
     +0.219 \vert 5 i_{13/2} \frac{3}{2} \rangle $ \\              \hline
 {} & {} & {} & {} & {} & $\vert 19 \rangle \; \; 0.995 \vert 4g_{7/2}
      \frac{7}{2} \rangle $ \\
 {} & {} & {} & {} & {} & $\vert 20 \rangle \; \; 0.967 \vert 4d_{5/2}
  \frac{5}{2} \rangle + 0.217 \vert 4 g_{7/2} \frac{5}{2} \rangle
  + 0.131 \vert 4 g_{9/2}\frac{5}{2} \rangle $ \\
 band 6 & 0.267 & 0.265 & 0 & 0 & $\vert 21 \rangle \; \;
     0.935 \vert 6i_{13/2} \frac{1}{2} \rangle + 0.334 \vert 6 g_{9/2}
     \frac{1}{2} \rangle $ \\
($[660]\frac{1}{2}^{+}$) & {} & {} & {} & {} & $\vert 22 \rangle
     \;\; 0.946 \vert 6i_{13/2}\frac{3}{2} \rangle + 0.303 \vert 6 g_{9/2}
      \frac{3}{2} \rangle $ \\
 {} & {} & {} & {} & {} & $\vert 23 \rangle \; \; 0.906 \vert 4d_{3/2}
      \frac{3}{2} \rangle + 0.245 \vert 4 s_{1/2} \frac{1}{2} \rangle
     +0.214 \vert 5 d_{5/2} \frac{3}{2} \rangle $ \\              \hline
\end{tabular}
\label{tab3}
\end{center}
\end{footnotesize}
\end{table}

\begin{figure}[ht]
\begin{center}
\includegraphics[scale=1.5,angle=0]{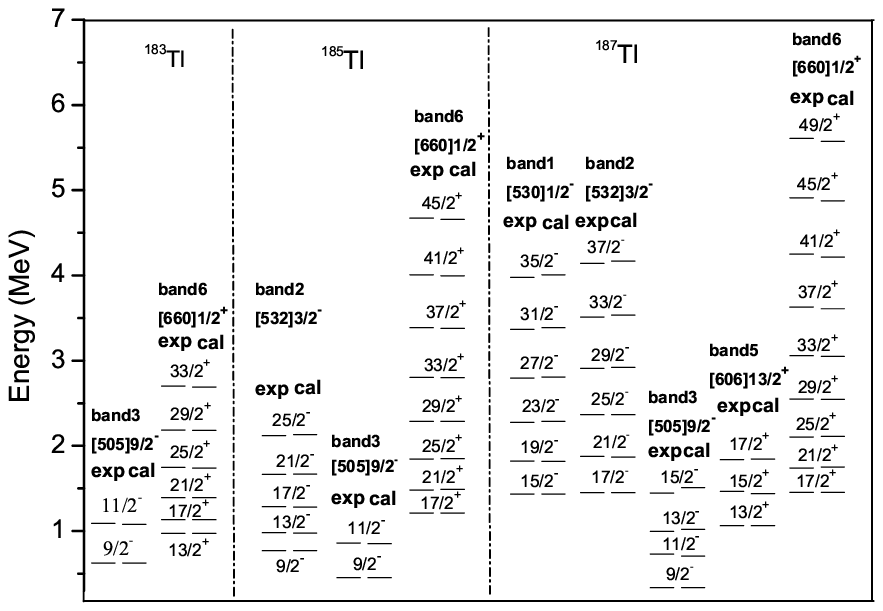}
\caption{Comparison of calculated energy levels of the rotational
bands in $^{183,185,187}$Tl with the experimental data (taken from
Refs.~\cite{Lane95,Muiku01}). }
\end{center}
\end{figure}

Because the experimentally observed rotational bands in $^{187}$Tl
are richer and more characteristic than those in $^{183}$Tl and
$^{185}$Tl, as a typical example, we analyze the band structure
and configuration in $^{187}$Tl in detail. To this end, we list
the total wavefunctions in terms of the single-particle orbitals
of the bands in $^{187}$Tl in Table 4.

\begin{table}[htpb]
\caption{The theoretically predicted main components of the
wavefunctions of the bands 1, 2, 3, 5 and 6 in $^{187}$Tl in terms
of the single-particle levels }
\begin{center}
\begin{tabular}{|c|l|}
\hline band & $I^{\pi}$ wavefunction $\vert \nu K \rangle $  \\
\hline
 {} & $\frac{15}{2}^{-} \;\; - 0.981\vert 22 \, \frac{1}{2} \rangle
                         + 0.189 \vert 19 \, \frac{1}{2} \rangle $ \\
 { }   & $\frac{19}{2}^{-} \;\; - 0.981\vert 22 \, \frac{1}{2}\rangle
                             + 0.190 \vert 19 \, \frac{1}{2} \rangle $ \\
band 1 & $\frac{23}{2}^{-}\;\; - 0.980\vert 22
              \, \frac{1}{2} \rangle + 0.192 \vert 19 \, \frac{1}{2} \rangle $ \\
($[530]\frac{1}{2}^{-}$) & $\frac{27}{2}^{-} \;\; - 0.980\vert 22
               \, \frac{1}{2}\rangle + 0.194 \vert 19 \, \frac{1}{2} \rangle $      \\
 { }   & $\frac{31}{2}^{-} \;\; 0.979\vert 22 \, \frac{1}{2} \rangle
                             - 0.196\vert 19 \, \frac{1}{2} \rangle $ \\
 { }   & $\frac{35}{2}^{-} \;\; 0.979\vert 22 \, \frac{1}{2} \rangle
                             - 0.197\vert 19 \, \frac{1}{2} \rangle $       \\
\hline
  { }   & $\frac{17}{2}^{- }\;\; - 0.964\vert 20 \, \frac{3}{2} \rangle
                            + 0.263 \vert 19 \, \frac{1}{2} \rangle $ \\
  { }   & $\frac{21}{2}^{-}\;\; 0.955\vert 20 \, \frac{3}{2} \rangle
                            - 0.294\vert 19 \, \frac{1}{2} \rangle $  \\
band 2  & $\frac{25}{2}^{-} \;\; 0.947\vert 20 \,
                  \frac{3}{2} \rangle - 0.321\vert 19 \, \frac{1}{2} \rangle $ \\
($[532]\frac{3}{2}^{-}$) & $\frac{29}{2}^{-} \;\; - 0.939\vert 20
              \, \frac{3}{2} \rangle + 0.343\vert 19 \, \frac{1}{2} \rangle $  \\
  { }   & $\frac{33}{2}^{-}\;\; - 0.932\vert 20 \, \frac{3}{2} \rangle
                            + 0.362\vert 19 \, \frac{1}{2} \rangle $   \\
  { }   & $\frac{37}{2}^{-}\;\; - 0.925\vert 20 \, \frac{3}{2} \rangle
                            + 0.378\vert 19 \, \frac{1}{2} > $         \\
\hline
   { }    & $\frac{9}{2}^{-} \;\;\; 0.815\vert 20 \, \frac{9}{2} \rangle
                            + 0.444\vert 21 \, \frac{7}{2} \rangle
                            + 0.231\vert 23 \, \frac{5}{2} \rangle $   \\
band 3    & $\frac{11}{2}^{-} \;\; 0.728\vert 20 \,
                \frac{9}{2} \rangle + 0.518\vert 21 \, \frac{7}{2} \rangle
                            - 0.299\vert 23 \, \frac{5}{2} \rangle $   \\
($[505]\frac{9}{2}^{-}$) & $\frac{13}{2}^{-}\;\; 0.500\vert 20 \,
                \frac{9}{2} \rangle + 0.479\vert 21 \, \frac{7}{2} \rangle
                            + 0.399\vert 23 \, \frac{5}{2} \rangle $  \\
   { }    & $\frac{15}{2}^{-}\;\; -0.584\vert 20 \, \frac{9}{2} \rangle
                            + 0.544\vert 21 \, \frac{7}{2} \rangle
                            - 0.388\vert 23 \, \frac{5}{2} \rangle $   \\
\hline

band 5 & $\frac{13}{2}^{+} \;\; 0.720\vert 23 \, \frac{13}{2}
                   \rangle  + 0.520 \vert 24 \, \frac{11}{2} \rangle
                            + 0.350 \vert 25 \, \frac{9}{2} \rangle $ \\
($[606]\frac{13}{2}^{+}$) & $\frac{15}{2}^{+} \;\; 0.580 \vert 23
              \, \frac{13}{2} \rangle - 0.560\vert 24 \, \frac{11}{2} \rangle
                              + 0.450\vert 25 \, \frac{9}{2} \rangle $ \\
   { }     & $\frac{17}{2}^{+} \;\; - 0.620\vert 23 \, \frac{13}{2} \rangle
                            - 0.390\vert 27 \, \frac{5}{2} \rangle
                            + 0.390 \vert 26 \, \frac{7}{2} \rangle $  \\
\hline

   {  }  & $\frac{17}{2}^{+} \;\; - 0.952\vert 21 \, \frac{1}{2} \rangle
                             - 0.298\vert 22 \, \frac{3}{2} \rangle $  \\
   {  }  & $\frac{21}{2}^{+} \;\; - 0.943\vert 21 \, \frac{1}{2} \rangle
                             - 0.326\vert 22 \, \frac{3}{2} \rangle $  \\
band 6   & $\frac{25}{2}^{+} \;\; 0.936\vert 21 \,
                 \frac{1}{2} \rangle + 0.347\vert 22 \, \frac{3}{2} \rangle $  \\
($[660]\frac{1}{2}^{+}$) & $\frac{29}{2}^{+}\;\; 0.930\vert 21 \,
                 \frac{1}{2} \rangle + 0.363\vert 22 \, \frac{3}{2} \rangle $  \\
   {  }  & $\frac{33}{2}^{+} \;\; - 0.924\vert 21 \, \frac{1}{2} \rangle
                             - 0.377\vert 22 \, \frac{3}{2} \rangle $  \\
   {  }  & $\frac{37}{2}^{+} \;\; 0.919\vert 21 \, \frac{1}{2} \rangle
                             + 0.388\vert 22 \, \frac{3}{2} \rangle $  \\
   {  }  & $\frac{41}{2}^{+} \;\; 0.915\vert 21 \, \frac{1}{2} \rangle
                             + 0.398\vert 22 \, \frac{3}{2} \rangle $  \\
\hline
\end{tabular}
\label{tab4}
\end{center}
\end{table}

From Table 4, we can recognize that the band 1 originates near
purely from the 22nd single particle orbital coupling with the
prolate even-even $^{186}$Hg core. Seen from Table 3, the 22nd
orbital contains mixing of 54.9{\%} $\vert 5 f_{7/2} \frac{1}{2}
\rangle $, 23.0{\%} $\vert 5 p_{3/2} \frac{1}{2} \rangle $ and
10.7{\%} $\vert 5 h_{9/2} \frac{1}{2} \rangle $ configurations.
Since the largest component is $\vert 5 f_{7/2} \frac{1}{2}
\rangle$, the band 1 can be assigned as the one arising mainly from
the configuration $[530]{\frac{1}{2}}^{-}$ ($\pi \, f_{7/2})$.
Meanwhile, from Table 4, we can see that the band 2 consists of
mixing of about 93{\%} 20th and 7{\%} 19th orbitals. Seen from Table
3, the 20th orbital contains 84.3{\%} of $\vert 5 h_{9/2}
\frac{3}{2}\rangle$ configuration. Thus, we can infer that the band
2 originates from the $[532]{\frac{3}{2}}^{-}$ configuration.
Similarly, combining Table 4 with Table 3, we can recognize that the
band 5, 6 originates mainly from the 23rd, 21st single particle
orbital, respectively, the 21st orbital contains 87.4{\%} of $\vert
6i_{13/2} \frac{1}{2} \rangle$ configuration, and the 23rd consists
of almost purely the $\vert 6 i_{13/2} {\frac{13}{2}} \rangle $
configuration. Therefore, the bands 5, 6 are based on the
configuration $\pi[606]{\frac{13}{2}}^{+}$, $\pi [660]
{\frac{1}{2}}^{+}$, respectively. The above calculated results of
the bands 1, 2, 5 and 6 are consistent with the prediction of
Ref.~\cite{Lane95}. And the configuration assignment of the bands 1
and 2 is a corroboration of that in Ref.~\cite{Lane95}. Furthermore,
the $[530]{\frac{1}{2}}^{-}$ ($f_{7/2}$) band involves about $10\%$
$h_{9/2}$ configuration and the $[532]{\frac{3}{2}}^{-}$ ($h_{9/2}$)
band involves only about $4\%$ $f_{7/2}$ configuration. On the other
hand, similar results are obtained for the corresponding bands in
$^{183,185}$Tl.

\begin{figure}[ht]
\begin{center}
\includegraphics[scale=1.0,angle=0]{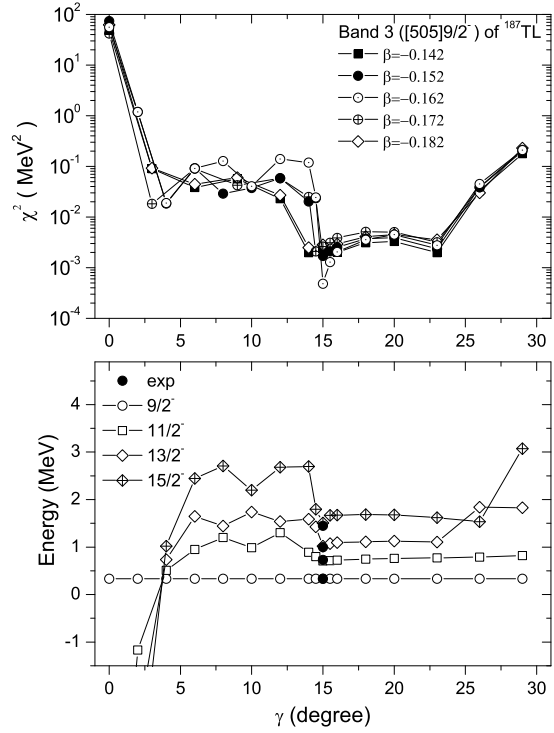}
\caption{Upper panel: calculation error $\chi^{2}$ against the
deformation parameter $\gamma$ at several axial deformation
parameter $\beta$'s of the band 3 in $^{187}$Tl. Lower panel:
variation of the calculated energy spectrum of the band 3
in $^{187}$Tl with fixed $\beta = -0.162$ against the value of
$\gamma$ and comparison with experimental data.
The experimental data  are taken from Ref.\cite{Lane95}. }
\end{center}
\end{figure}

\begin{figure}[ht]
\begin{center}
\includegraphics[scale=1.0,angle=0]{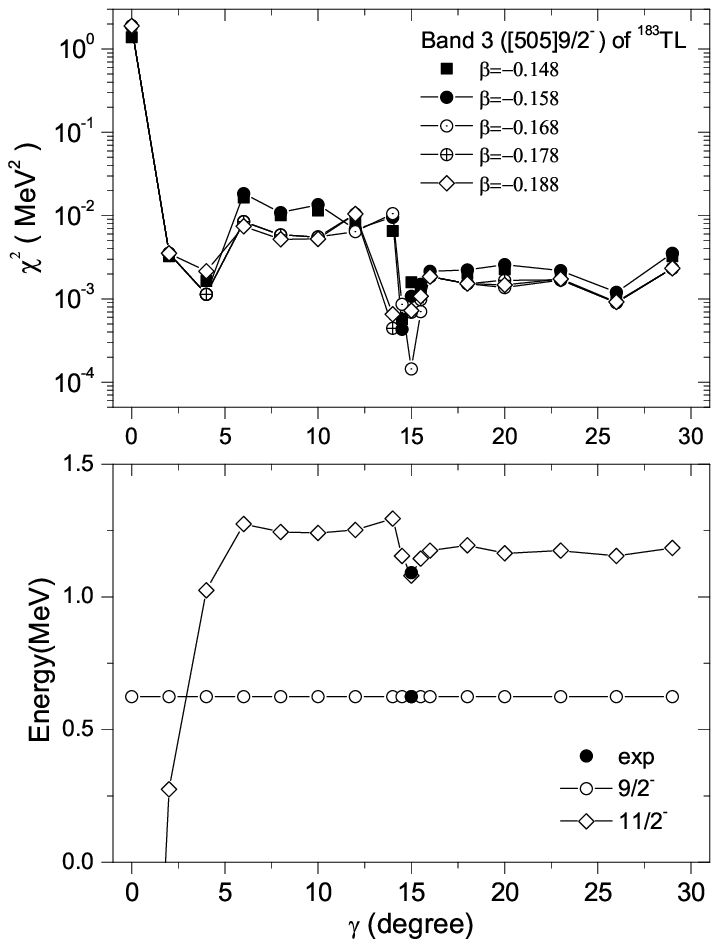}
\caption{Upper panel: calculation error $\chi^{2}$ against the
deformation parameter $\gamma$ at several axial deformation
parameter $\beta$'s of the band 3 in $^{183}$Tl. Lower panel:
variation of the calculated energy spectrum of the band 3
in $^{183}$Tl with fixed $\beta = -0.168$ against the value of
$\gamma$ and comparison with experimental data.
The experimental data  are taken from Ref.\cite{Muiku01}. }
\end{center}
\end{figure}

\begin{figure}[htbp]
\begin{center}
\includegraphics[scale=1.0,angle=0]{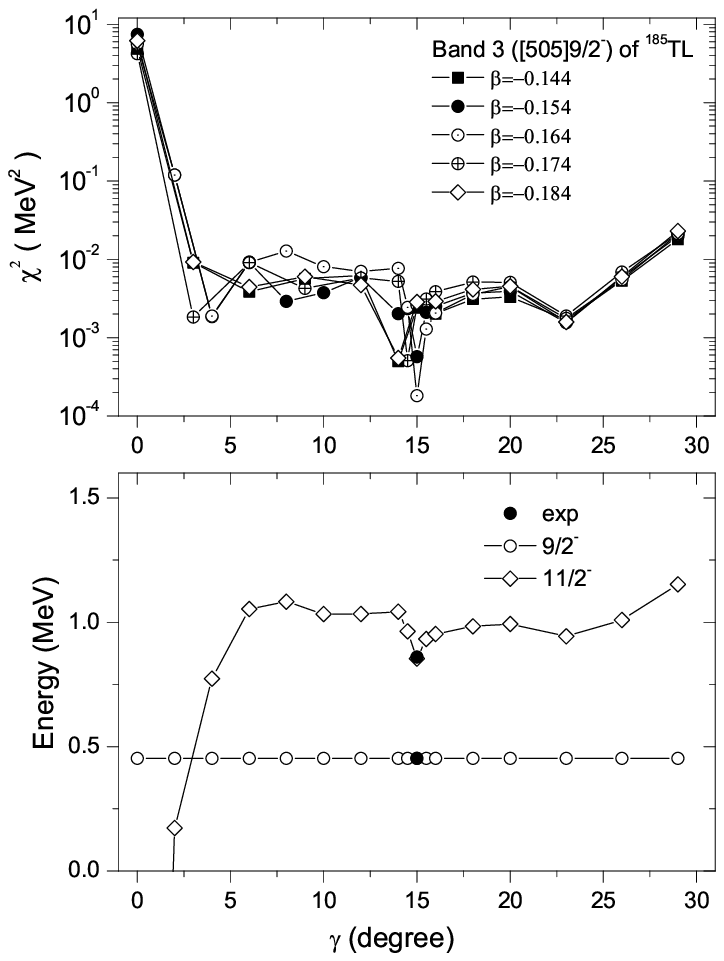}
\caption{Upper panel: calculation error $\chi^{2}$ against the
deformation parameter $\gamma$ at several axial deformation
parameter $\beta$'s of the band 3 in $^{185}$Tl. Lower panel:
variation of the calculated energy spectrum of the band 3 in
$^{185}$Tl with fixed $\beta = -0.164$ against the value of
$\gamma$ and comparison with experimental data. The experimental
data  are taken from Ref.~\cite{Lane95}. }
\end{center}
\end{figure}

\begin{figure}[htbp]
\begin{center}
\includegraphics[scale=1.0,angle=0]{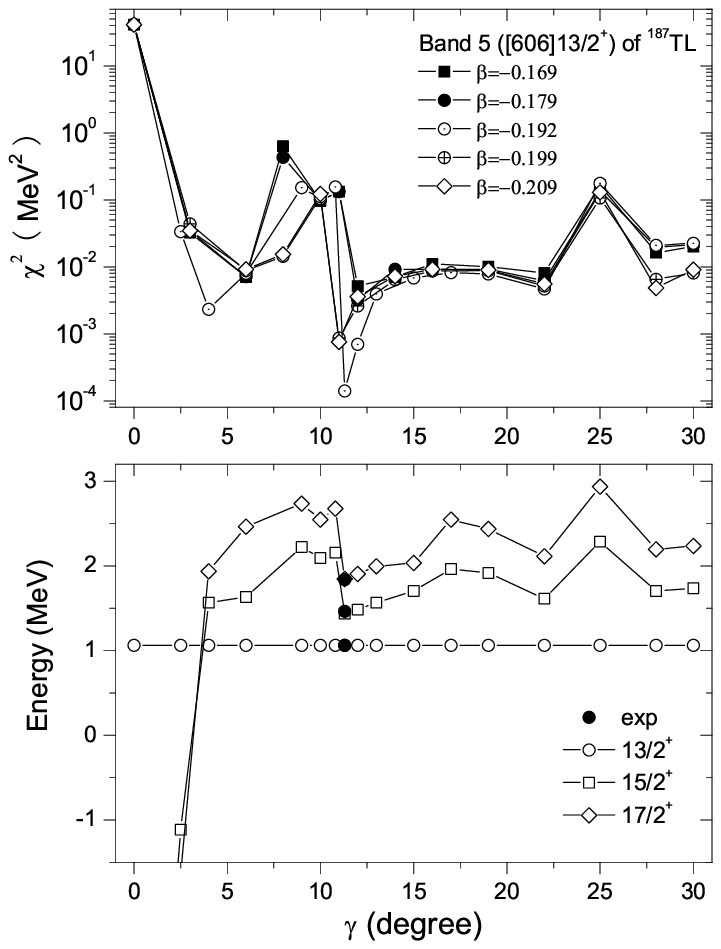}
\caption{Upper panel: calculation error $\chi^{2}$ against the
deformation parameter $\gamma$ at several axial deformation
parameter $\beta$'s of the band 5 in $^{187}$Tl. Lower panel:
variation of the calculated energy spectrum of the band 5 in
$^{187}$Tl with fixed $\beta = -0.192$ against the value of
$\gamma$ and comparison with experimental data. The experimental
data  are taken from Ref.~\cite{Lane95}. }
\end{center}
\end{figure}

In order to investigate the deformation nature of band 3 in
$^{183,185,187}$Tl, we illustrate at first the calculation error
$\chi^2 = \frac{1}{N} \sum_{j} (E^{cal.}_{j} - E^{exp.}_{j} )^2 $
of the energy spectrum of band 3 in $^{187}$Tl (where $N$ is the
number of levels in the band) with respect to the value of
$\gamma$ at several $\beta$'s in the upper panel of Fig.~2. We
also display the variation of the calculated energy spectrum
against the value of $\gamma$ at the best fitted $\beta \,
(-0.162)$ and the comparison with experimental data of the band in
the lower panel of Fig.~2. The upper panel of Fig.~2 shows that
the variation of the axial deformation parameter $\beta$ (except
for that with angular deformation parameter $\gamma$ in special
region) does not affect the calculation error $\chi ^{2}$ so
drastically as that of the $\gamma$ does. Combining the upper
panel and the lower panel of Fig.~2, one can notice clearly that,
for zero $\gamma$, the calculation error $\chi ^{2}$ is quite
large (about 60) and the calculated level sequence is not
consistent with experiments. As the $\gamma$ increases to 3-5
degrees, the calculated level sequence becomes consistent with the
experimental one and the $\chi ^{2}$ decreases to about $10^{-1}$.
For the value of $\gamma$ in the region 3 to 14 degrees, the
calculation error $\chi ^{2}$ maintains around $10^{-1}$. When
$\gamma = 15 ^{\circ}$, the $\chi^{2}$ with $\beta = - 0.162$
becomes suddenly the minimum ($\sim 10^{-4}$) of the $\chi^{2}
(\beta , \gamma)$ and the calculated energy spectrum agrees with
experimental data very well. As $\gamma$ increases from 15 degrees
further, the $\chi^{2}$ increases to around $10 ^{-3}$, even to
$10^{-1}$. Moreover, in the case of $\beta = -0.162$, even though
the calculated energies of the states with lower angular momentum
do not deviate from experimental data obviously, the ones with
higher angular momentum do drastically. It is then evident that,
when the deformation parameters $(\beta,\gamma) = (-0.162,
15^{\circ})$, the calculated energy spectrum agrees best with
experimental data. It indicates that the band 3 of $^{187}$Tl is
in triaxial oblate deformation. In addition, from Table 4, we
notice that the band 3 in $^{187}$Tl originates mainly form the
20th single particle orbital. As can be seen from inspecting Table
3, the 20th orbital contains 97.2{\%} of $\vert 5 h_{9/2}
\frac{9}{2} \rangle$ configuration. Therefore, the band 3 can be
identified as the one arising from the proton configuration
$[505]{\frac{9}{2}}^{-}$ ({$\pi \,h_{9/2})$ coupled to a triaxial
oblate deformed core. It provides then a corroboration of the
conjecture in Ref.~\cite{Lane95}. Similar results for the bands 3
in $^{183,185}$Tl are obtained, too (the calculation errors
$\chi^{2}$ of the energy separations against the value of $\gamma$
at several $\beta$'s and the comparison of the calculated energy
spectrum with $\gamma \in (0^{\circ}, 29^{\circ})$ and $\beta =
-0.168$ ($-0.164$) with experimental data are illustrated in
Fig.~3 (4) for $^{183}$Tl ($^{185}$Tl) ). The deformation
parameters can then be fixed as $(-0.168, 15^{\circ})$, $(-0.164,
15^{\circ})$ for the band~3 of $^{183}$Tl, $^{185}$Tl,
respectively. These results confirm the assumption that the band
originated from orbital $[505]{\frac{9}{2}}^{-}$ ($\pi \,
h_{9/2})$ may be in triaxial oblate deformation\cite{Lane95}.
Besides, from the calculation error $\chi ^{2}$ and the comparison
between the obtained energy spectrum with $\beta = -0.192$ and the
experimental data of the band $[606]\frac{13}{2}^{+}$ shown in
Fig.~5, one can recognize that the band~5 ($[606]\frac{13}{2}^{+}$
band) is also a triaxial oblate deformation band (with $(\beta ,
\gamma ) = (-0.192, 11.3^{\circ})$ ).

From the calculations one may recognize that the deformation parameter
$\gamma$ influences the results more drastically than the axial
deformation parameter $\beta$. Analyzing the obtained single
particle configuration we know that the states in the bands
$[505]\frac{9}{2}^{-}$ and $[606]\frac{13}{2}^{+}$ involve much more
complicated single particle Nilsson configurations than the other
bands. Such a result is consistent with that once given for nucleus
$^{127}$I in Ref.~\cite{SLZ04}. We infer then that the reason for
the $\gamma$-degree of freedom to play more important role than
$\beta$ may be that it induces more obvious mixing among the single
particle Nilsson configurations.

\section{Conclusion and Remarks}

In summary, we have systemically calculated the energy spectra, the
deformations and wavefunctions of the rotational bands in nuclei
$^{183,185,187}$Tl in the particle triaxial-rotor model with
variable moment of inertia. The calculated energy spectra of the
bands agree quite well with the experimental data. The configuration
of the bands in $^{187}$Tl is analyzed in detail as an example.
Meanwhile we have also calculated the variation of the configuration
of single-particle levels against the deformation parameters $\beta$
and $\gamma $. Considering both the parameters fitted and the
agreement between calculated results and experimental data, we
conclude that the rotation-aligned band structures observed in
$^{183,185,187}$Tl are due to one of the $[530]{\frac{1}{2}}^{-}$,
$[532]{\frac{3}{2}}^{-}$, $[660]{\frac{1}{2}}^{+}$ proton
configurations coupled to a prolate deformed core. Meanwhile, the
negative parity bands built upon the ${\frac{9}{2}}^{-}$ isomeric
states in $^{183,185,187}$Tl are formed by a proton with the
$[505]{\frac{9}{2}}^{-}$ configuration coupled to a core with
triaxial oblate deformation $(\beta , \gamma) = (-0.168,
15^{\circ}), \, (-0.164, 15^{\circ}), \, (-0.162, 15^{\circ})$,
respectively, and the positive parity band on the
${\frac{13}{2}}^{+}$ isomeric state in $^{187}$Tl is generated by a
proton with configuration $[606]{\frac{13}{2}}^{+}$ coupled to a
triaxial oblate core with deformation parameters $(\beta , \gamma) =
(-0.192 , 11.3^{\circ})$. In short, the nuclei $^{183,185,187}$Tl
involve quite rich shape coexistence.
Meanwhile our present calculation provides a clue that the triaxial
deformation may arise from the mixing of single particle Nilsson
configurations. To understand it much better, more investigations
are required.

\bigskip

This work was supported partially by the Natural Science Foundation
of Guangdong Province with contract No. 04011642, partially by the
Natural Science Research Foundation of the Education Department of
Guangdong Province with contract No. Z02069, and partially by the
National Natural Science Foundation of China with contract Nos.
10425521, 10135030 and 10075002. One of the authors (YXL) thanks
also the support by the Major State Basic Research Development
Program under Grant No. G2000077400, the Key Grant Project of
Chinese Ministry of Education (CMOE) under contract No.~305001, the
Foundation for University Key Teacher by the CMOE and the Research
Fund for the Doctoral Programme of Higher Education of China with
grant No. 20040001010.


\end{document}